\documentclass[12pt]{article}
\usepackage{amsfonts}
\usepackage{latexsym}
\usepackage{amsmath}
\usepackage{amssymb}
\usepackage{amssymb}

\newcommand{\newsection}{    
\setcounter{equation}{0}\section}
\def\appendix#1{\addtocounter{section}{1}\setcounter{equation}{0}
\renewcommand{\thesection}{\Alph{section}}
\section*{Appendix \thesection\protect\indent \parbox[t]{11.15cm}{#1}}
\addcontentsline{toc}{section}{Appendix \thesection\ \ \ #1}}

\newcommand{\be}{\begin{eqnarray}}
\newcommand{\ee}{\end{eqnarray}}
\newcommand{\bea}{\begin{eqnarray}}
\newcommand{\eea}{\end{eqnarray}}
\newcommand{\ba}{\begin{array}}
\newcommand{\ea}{\end{array}}

\newenvironment{frcseries}{\fontfamily{frc}\selectfont}{}

\def \la {\label}

\def\a{\alpha}
\def\b{\beta}

\def\g{\gamma}

\font\mybb=msbm10 at 11pt

\def\bb#1{\hbox{\mybb#1}}

\def\bR {\bb{R}}

\begin{document}
\begin{titlepage}
\begin{center}
\vspace*{-1.0cm}

\vspace{2.0cm}
{\Large \bf  Killing-Yano equations with torsion,  worldline actions and G-structures} \\[.2cm]

\vspace{1.5cm}
 {\large G.~ Papadopoulos}

\vspace{0.5cm}

Department of Mathematics\\
King's College London\\
Strand Street\\
London WC2R 2LS, UK\\
\small{george.papadopoulos@kcl.ac.uk}
\vspace{0.5cm}

\end{center}

\vskip 1.5 cm
\begin{abstract}
We determine the geometry of the target spaces of  supersymmetric
non-relativistic particles with  torsion and  magnetic couplings,
and with symmetries generated by the fundamental forms of G-structures
for $G= U(n), SU(n), Sp(n), Sp(n)\cdot Sp(1), G_2$ and $Spin(7)$. We
find that the Killing-Yano  equation, which arises as a condition for the invariance
of the worldline action, does not always determine
the torsion coupling uniquely in terms of the metric and fundamental forms.
We show that there are several connections with skew-symmetric torsion for  $G=U(n), SU(n)$ and $G_2$ that
solve the invariance conditions. We  describe  all these compatible connections
 for each of the $G$-structures and explain the geometric
nature of the couplings.

\end{abstract}

\end{titlepage}

\setcounter{section}{0}
\setcounter{subsection}{0}



\section{Introduction}

It is well known that worldvolume actions of particles and strings
admit $W$-type of symmetries generated by spacetime forms \cite{phgp}.
In string theory,  such forms are parallel with respect to a metric connection.  So such symmetries exist provided that
the spacetime is a manifold with special holonomy and therefore special G-structure. These  symmetries are part of the chiral
W-algebra of the worldvolume theory \cite{vafa, deBoer} and so they are instrumental
in the investigation of quantum theory.

On the other hand for particles, the spacetime forms that generate the symmetries are not always parallel. Instead they
satisfy the Killing-Yano (KY) equation with respect to a metric connection\footnote{The connection depends on the choice of
couplings in the action.}. This is a weaker condition than that which arises for strings, and first observed
in the construction of worldline actions with more than one supersymmetries \cite{gibbons, sfetsos, gbksgp}.
The existence of such symmetries  does not necessarily imply the reduction
of either the holonomy of the connection or the G-structure of spacetime. Nevertheless
a large class of examples can be found by making  an identification of the forms that generate the symmetries with the fundamental
forms of a G-structure.  Such an analysis
has been done for a particle action with only a metric coupling in \cite{gpp}. In such case, the KY equations are respect to
the Levi-Civita connection and the structure groups considered are
 \bea
U(n)~(2n), SU(n)~(2n), Sp(n)~(4n), Sp(n)\cdot Sp(1)~(4n), G_2~(7), Spin(7)~(8)~,
\la{gst}
\eea
 where
in parenthesis is the dimension of the associated manifold\footnote{We assume that the spacetime is a product $\bR^{k,1}\times M$ and we shall focus
on the particle dynamics on the Riemannian manifold $M$.} $M$. It has been found that
apart from the $U(n)$, $SU(n)$ and $G_2$ cases, the KY equations imply that
the fundamental forms are parallel and so the holonomy group of the Levi-Civita
connection reduces to $G$. For the $U(n)$, $SU(n)$ and $G_2$  G-structures, the KY equations do not always imply that the fundamental forms are parallel but
rather restrict the G-structure of the spacetime, ie
some of the Gray-Hervella type of classes \cite{gray}, or a linear combination of them,  must vanish.
An extension of these results to other G-structures from those stated in (\ref{gst}) has been given in \cite{santillan}.

In this paper, we shall extend the analysis of the symmetries  to particle actions which apart from the metric $g$  also have a 3-form torsion  $c$ and magnetic $A$ couplings. The presence of torsion modifies the conditions that are required  for  the invariance of the action under symmetries generated by spacetime forms. In particular, one finds that
\bea
\hat\nabla_{j_1} L_{j_2\dots j_{k+1}}=\hat\nabla_{[j_1} L_{j_2\dots j_{k+1}]}
\la{eqn1}
\eea
and
\bea
(-1)^{k-1}\partial_{[j_1} \big(L^m{}_{j_2\dots j_{k}} c_{j_{k+1}j_{k+2}]m}\big)-
{(k+1)\over6} L^m{}_{[j_1\dots j_{k-1}} dc_{j_{k} j_{k+1} j_{k+2}]m}=0\ ,
\la{eqn2}
\eea
where $L$ is the form that generates the symmetries and
$\hat\nabla$ is the metric connection with torsion $c$. The magnetic coupling $A$
does not restrict the geometry but there is a condition on $dA$ which will appear in section 2. The first condition (\ref{eqn1})
is the KY equation with respect to the connection $\hat\nabla$.
This generalizes the standard KY equation which is taken with respect to
 the Levi-Civita connection \cite{yano}, and has found  applications
 in the integrability of geodesic flows and Klein-Gordon and Dirac equations on curved manifolds \cite{penrose}-\cite{page}.  In a similar context, the KY equation with respect to $\hat\nabla$ has also been considered before in \cite{warnick1, warnick2}.  The second
condition is an additional restriction on $c$ which does not arise for particle systems with just a metric coupling.

The main aim of this paper is to solve both invariance conditions (\ref{eqn1}), (\ref{eqn2})
assuming that the forms $L$ that generate the symmetries are the fundamental forms
 of the $G$-structures\footnote{There are many other G-structures that can be considered
 but the focus will be on those given in (\ref{gst}) as they are related to irreducible Riemannian
manifolds, see eg \cite{santillan}.} stated in (\ref{gst}).   We shall find that for the G-structures $U(n)$, $SU(n)$ and $G_2$  and for the symmetries generated by some of the fundamental forms, there are several
 connections with skew-symmetric torsion that solve both  (\ref{eqn1})
and (\ref{eqn2}). As a result, the couplings of the particle action are not uniquely
determined in terms of the metric and the fundamental forms. This is unlike the string case where all the worldvolume
couplings  are given in terms of the metric and fundamental forms of the G-structures
(\ref{gst}) that generate the symmetries. We identify all connections which are compatible with both
invariance conditions (\ref{eqn1}) and (\ref{eqn2}). In the remaining cases, we show that the KY equation implies that
the fundamental forms are parallel with respect to the connection with skew-symmetric
torsion $\hat\nabla$. Such a connection is unique and so all the couplings
of the action are determined in terms of the metric and fundamental forms of the
G-structure.

Furthermore, we identify the restrictions on the G-structures
such that KY equation (\ref{eqn1}) admits a solution. Typically, these
are expressed as the vanishing conditions of  some of the Gray-Hervella type of classes, or a linear
combination of them. We show that in all cases, the restrictions on the Gray-Hervella type
of classes are the same as those required for $M$ to admit a connection $\hat{\mathring \nabla}$ with
skew-symmetric torsion compatible with the $G$-structure.
Therefore on all manifolds that (\ref{eqn1}) has
a solution, there is a connection with skew-symmetric
torsion $\hat{\mathring \nabla}$, which may be different from $\hat\nabla$,
   such that $\hat{\mathring \nabla}L=0$ for $L$ fundamental forms of a $G$-structure.

This paper is organized as follows. In section two, we present a derivation
of the invariance conditions (\ref{eqn1}) and (\ref{eqn2}) as well as a refinement
for the special case that $\hat\nabla$ is compatible with the corresponding G-structure.
In section three, we present the solution of (\ref{eqn1}) and (\ref{eqn2}) for the
$U(n)$ structure. In section four, we solve the invariance conditions for the
$SU(n)$ structure, and similarly in section five for the $Sp(n)$ and $Sp(n)\cdot Sp(1)$
structures. The analysis of the $G_2$ and $Spin(7)$ cases are given in sections
six and seven, respectively.

\section{Particle actions and their symmetries}

\subsection{Invariance of action}

Consider  particle with $N=1$ supersymmetry propagating in a manifold $M$. The particle positions are maps
$X$ from $\Xi^{1\vert1}$ superspace into $M$. The most general action  with up to two derivative terms
written in superspace
can be written \cite{cp} as
\bea
I=-{i\over2}\int dt\, d\theta~\big(~ g_{ij} DX^i \partial_tX^j-{i\over6} c_{ijk} DX^i DX^j DX^k+ A_i DX^i ~\big)~,
\la{actn1}
\eea
where $g$ and $c$ is a metric and a 3-form  in $M$, respectively, while the magnetic gauge potential $A$ is a locally defined 1-form on $M$.
Moreover
\bea
D^2=i\partial_t~, \quad D=\partial_\theta+i\theta\partial_t\ ,
\eea
where $(t, \theta)$ are the coordinates of $\Xi^{1\vert1}$.

Suppose that this action is invariant under the transformation
\bea
\delta X^i=L^i{}_{j_1\dots j_{k-1}} DX^{j_1}\dots DX^{j_{k-1}}~,
\la{sym}
\eea
where $L$ is a k-form on $M$. In such a case, the conditions for invariance
of the action are given\footnote{In our conventions, $\hat\nabla_i X_j\equiv\nabla_i X_j-\frac{1}{2}c^k{}_{ij}{}X_k\ .$} in (\ref{eqn1}), (\ref{eqn2}) and
\bea
dA_{i[j_1} L^i{}_{j_2\dots j_k]}=0~.
\la{eqn3}
\eea
Observe that (\ref{eqn2}) can
be rewritten as
\bea
{\cal L}_L c+ (-1)^{k-1} {k-1\over 2}  i_L dc=0~,
\la{feqn2}
\eea
where ${\cal L}_L=d i_L+ (-1)^{k-1} i_L d$ is the Lie derivative with respect to $L$, see eg \cite{phgp}. Therefore $c$ is invariant provided that either $k=1$ or $dc=0$.
Similarly, we have that (\ref{eqn3}) can be expressed as
\bea
i_L dA=0~.
\la{feqn3}
\eea

The invariance conditions (\ref{eqn1}) and (\ref{eqn2}) for the action are different from those that have appeared in 2-dimensional sigma
 models related to the propagation of strings in curved spaces, see \cite{phgp, hps}. In particular, the invariance
condition in 2-dimensional models implies that $L$ is parallel with respect $\hat\nabla$ together with $dc=0$. The
commutators of  symmetries (\ref{sym}) are similar to those  investigated in \cite{phgp, hps}
for similar transformations in (1,0)-supersymmetric 2-dimensional sigma models
and they will not be repeated here.

Before we proceed to investigate the  conditions (\ref{eqn1}) and (\ref{eqn2}) for various G-structures, let us consider the special case where
\bea
\hat\nabla_i L_{j_1\dots j_n}=0~.
\la{neqn1}
\eea
Clearly this condition implies (\ref{eqn1}). Moreover it simplifies (\ref{eqn2}). In particular using the integrability
condition of (\ref{neqn1}) and a Bianchi identity, one finds that
 \bea
 {\cal L}_L c=0~,
 \eea
 and  so (\ref{eqn2}) can be re-expressed as
 \bea
 i_L dc=0~,
 \la{neqn2}
 \eea
 or equivalently in components
\bea
L^m{}_{[j_1\dots j_{k-1}} dc_{j_{k} j_{k+1} j_{k+2}]m}=0~.
\eea
It turns out that for many G-structures, the KY equation (\ref{eqn1}) implies (\ref{neqn1}). In such cases, it is simpler
to consider (\ref{neqn2}) rather than (\ref{eqn2}).

\subsection{1-form symmetries}

The simplest case to consider  is that for which the symmetry is generated by a 1-form $X$.
 Then (\ref{eqn1}) implies that the associated vector field to $X$ is Killing. Moreover as $k=1$, (\ref{eqn2}) implies that  $c$ is invariant under the action of $X$.
Therefore considering also (\ref{eqn3}),  one has
\bea
\nabla_{(i} X_{j)}=0~,~~~{\cal L}_X c=0~,~~~i_X dA=0~.
\la{kyk}
\eea
Thus there are such symmetries provided that $M$ admits a Killing vector field
and $c$ is invariant under the action of isometries.

It is instructive to compare these conditions with those that arise by taking
$\hat\nabla X=0$ as in (\ref{neqn1}). In this case, one finds that
 (\ref{neqn1}),  (\ref{neqn2}) and (\ref{eqn3})  imply that
\bea
\nabla_{(i} X_{j)}=0~,~~~dX=i_Xc~,~~~i_X dc=0~,~~~i_X dA=0
\la{pk}
\eea
Clearly (\ref{pk}) implies (\ref{kyk}) but not conversely.

\subsection{Solution of the invariance equations for general fundamental  forms}

The solution of  (\ref{eqn1}),  (\ref{eqn2}) and (\ref{eqn3}) is simplified  whenever $L$ is
identified with a fundamental form of a $G$-structure. This is because all
such forms are invariant under the action of $G$. To solve (\ref{eqn1}) observe that
 for all $G$-structures in (\ref{gst}),
the Lie algebra of $G$, $\mathfrak{g}$, is included in  the space of 2-forms, $\Lambda^2(F)$, of the typical fibre $F$ of
the tangent bundle of $M$. So, one can write
$\Lambda^2(F)=\mathfrak{g}\oplus \mathfrak{g}^\perp$. Using this decomposition, one
can decompose $\hat\nabla$ as
 \bea
 \hat\nabla= \pi(\hat\nabla)\oplus \sigma(\hat\nabla)~.
 \la{denabla}
 \eea
 Clearly, $\pi(\hat\nabla)L=0$ and so (\ref{eqn1}) turns into an equation for $\sigma(\hat\nabla)$ which is an element
 of $F\otimes \mathfrak{g}^\perp$. Note  that $\sigma(\hat\nabla)$ may not be a 3-form, though it will turn out to be the
 case for all G-structures that we shall examine.  This decomposition is identical
 to that used in \cite{gpp} to solve the KY equation associated with the
 Levi-Civita connection. One of the advantages of this observation is that the results
 of \cite{gpp} can be used to determine $\sigma(\hat\nabla)$ and a new calculation is not needed. However, the presence of torsion
 leads to weaker conditions on the G-structures required for the existence  of solutions to the KY equation (\ref{eqn1})
 than those we have found in \cite{gpp} for the existence of solutions to the KY
 equation associated with the Levi-Civita connection.

 To solve (\ref{eqn2}) and (\ref{eqn3}), one decomposes $\Lambda^4(F)$ and $\Lambda^2(F)$ in irreducible representations of $G$,
 respectively. The restrictions imposed
 by  (\ref{eqn2}) and (\ref{eqn3}) can be expressed as the vanishing conditions
 of some of these irreducible representations.

 In what follows, the conventions that we shall use for $G$-structures, included the choice
 of representatives for fundamental forms, are those
 given in \cite{uggp}.  Moreover, a collection of the expressions for the
 torsion $\mathring c$ of the compatible connections $\hat{\mathring c}$
 to the G-structures in (\ref{gst}) in terms of the metric and fundamental forms
 can also be found in \cite{uggp}.

\newsection{$U(n)$ structure}

The form that generates the symmetry is the Hermitian 2-form $\omega(X,Y)= g(X, IY)$ of an  almost complex structure $I$, where the
metric $g$ is  Hermitian with respect to $I$, ie $g(IX, IY)=g(X,Y)$.  There are two cases to consider
depending on whether the almost complex structure $I$
is integrable or not. In the integrable case, this symmetry is  identified
  with a second supersymmetry of the system. In the non-integrable case, the symmetry
  is again associated with a second supersymmetry but an additional charge should be
  included in the supersymmetry algebra generated by the Nijenhuis tensor \cite{delius}. First, we shall consider the integrable case and then
we shall extend our results to manifolds with non-integrable almost complex structures.

\subsection{Integrable complex structures}

\subsubsection{Solution of the conditions}

It is convenient to do the analysis of the conditions (\ref{eqn1}) and (\ref{eqn2}) in complex coordinates. For this, decompose $c$ with respect to the complex structure $I$ as
 \bea
 c=c^{3,0}+c^{2,1}+c^{1,2}+c^{0,3}~,
 \eea
 where $c^{1,2}$ and $c^{0,3}$  are complex conjugate to $c^{2,1}$ and $c^{3,0}$, respectively. Next decomposing (\ref{eqn1}) in (3,0) and (2,1) parts, one finds that
\bea
c^{2,1}=-i\partial \omega~,
\la{cc21}
\eea
where $\partial$ is the holomorphic exterior derivative. The (3,0) part of $c$ is not restricted by (\ref{eqn1}) and so it can be arbitrary.

Next repeating the procedure for (\ref{eqn2}), one finds that from the vanishing of
(4,0) component one gets
\bea
\partial c^{3,0}=0~.
\la{cc30}
\eea
The remaining components are identically zero.
Thus  the full content of both (\ref{eqn1}) and (\ref{eqn2}) equations  are the  conditions given in (\ref{cc21}) and (\ref{cc30}). The above calculation
has been made  in \cite{delduc} reaching a similar conclusion.
It remains to solve (\ref{eqn3}). This implies that $dA$ is a (1,1) form, ie
$dA^{2,0}=0$.

\subsubsection{Geometry}

It is clear from (\ref{cc21}) and (\ref{cc30}) the 3-form $c$ that appears in the action
is not uniquely determined in terms of the metric and the Hermitian form $\omega$. This is because the (3,0) component  can be any $\partial$-closed (3,0)-form on $M$.
A consequence of this is that {\it there is not a unique} connection $\hat\nabla$ that
solves the KY equation (\ref{eqn1}).

A special case is to take  $c^{3,0}=0$ and denote the remaining non-vanishing
  components with $\mathring c$. In such case, $\mathring c$ is entirely given in (\ref{cc21}) and it is uniquely expressed in terms of the metric and $\omega$. Moreover, $I$ is parallel with respect to the connection $\hat{\mathring \nabla}$ with skew-symmetric torsion $\mathring c$,  $\hat{\mathring \nabla}I=0$.
  $\hat{\mathring \nabla}$ coincides with the Bismut connection.

In conclusion, for  a fixed metric and complex structure the KY equation (\ref{eqn1}) admits many solutions on any Hermitian
manifold. These solutions are parameterized by (3,0)-forms on $M$. If
in addition (\ref{eqn2}) is also imposed, then the (3,0)-forms are restricted to be $\partial$-closed.

A physical consequence  of the above geometric results is that the 3-form coupling
of the action (\ref{actn1}) for models with two supersymmetries are not uniquely
determined in terms of the metric and complex structure.

\subsection{Non-integrable complex structures}

Next suppose that the almost complex structure $I$ is non-integrable.
To investigate the consequences of (\ref{eqn1}) and (\ref{eqn2}) on the geometry, it is convenient to introduce
a compatible frame to the $U(n)$ structure as
\bea
ds^2=2\delta_{\a\bar\b} e^\a e^{\bar\b}~,~~~\omega=-i \delta_{\a\bar\b} e^\a\wedge e^{\bar\b}~.
\eea
In this frame, the (3,0) component of KY equation (\ref{eqn1}) implies that
\bea
\Omega_{\a,\b\g}=\Omega_{[\a,\b\g]}~,
\la{ac1}
\eea
where $\Omega$ is the frame Levi-Civita connection.
This is a geometric restriction which is equivalent to requiring that the Nijenhuis
tensor of the almost complex structure is skew-symmetric in all three indices. This
is not always the case for every $2n$-dimensional almost Hermitian manifold.
In fact, it is required that one of the Gray-Hervella classes \cite{gray} must vanish, $W_2=0$.

Next the (2,1) part of (\ref{eqn1}) implies that
\bea
c_{\a\b\bar\g}=2\Omega_{\bar\g,\a\b}~.
\eea
This is not a geometric condition. It simply expresses some of the components of $c$
in terms of the metric and almost complex structure of $M$.

To proceed with solving (\ref{eqn2}), it is convenient to observe that
the geometric condition (\ref{ac1}) implies that there is another metric connection $\hat{\mathring \nabla}$ on $M$ with skew-symmetric
 torsion $\mathring c$ such that
\bea
 \hat{\mathring \nabla} \omega=0~.
 \la{unpar}
 \eea
 In particular, we have that
 \bea
 \mathring c_{\a\b\g}=2\Omega_{\a,\b\g}~,~~~\mathring c_{\bar\a\b\g}=2\Omega_{\bar\a,\b\g}~.~~~
 \eea
 In fact an explicit expression for $\mathring c$ in terms of the $I$ and $g$
 can be found in \cite{ivanov}, see also \cite{uggp}.
 Writing
 \bea
 \hat\nabla=\hat{\mathring \nabla}+ S~,
 \eea
 we observe that
 \bea
 S^{2,1}=0~,
 \la{uns}
 \eea
 and
 \bea
 c^{3,0}=\mathring c^{3,0}+ S^{3,0}~.
 \la{uuns}
 \eea
 This decomposition is the same as that in (\ref{denabla}) for $G=U(n)$
 with $\pi(\hat\nabla)= \hat{\mathring \nabla}$ and $\sigma(\hat\nabla)=S$.

\subsubsection{$S=0$}

In this case $c=\mathring c$. As a result, (\ref{eqn2}) can be rewritten as
(\ref{neqn2}). In turn this implies that
\bea
d\mathring c^{4,0}=d\mathring c^{3,1}=0~,
\la{conr}
\eea
but the (2,2) part of $d\mathring c$ remains unconstrained.

The Bianchi identities of $\hat R$ together with the integrability
condition of  (\ref{unpar}) allow for further conditions on $\mathring c$.
In particular, one finds that
\bea
c^m{}_{[\a_1\a_2} c_{\a_3\a_4]m}=0~,~~~\partial \mathring c^{1,2}=\bar\partial \mathring c^{2,1}~.
\eea
Moreover the second condition in (\ref{conr}) can be rewritten as
\bea
 \hat{\mathring \nabla}_{\bar\b} \mathring c_{\a_1\a_2\a_3}=0~.
 \la{amo2}
 \eea
Unlike the case with an integrable complex structure, it is not straightforward
to solve (\ref{conr}). Nevertheless, the conditions (\ref{conr}) can be
easily checked for particular examples of almost Hermitian manifolds.

\subsubsection{$S\not=0$}

The investigation of (\ref{eqn2}) can be organized in different ways. One way
is to observe that ${\cal L}_I$ coincides with the exterior derivative
with respect to $I$, $d_I$. Then using $d_I \mathring c=0$ and that $S$ is (3,0) and (0,3) form, we can write
(\ref{eqn2}) as
\bea
d_I S- {1\over 2}  i_I dS- {3\over 2}  i_I d\mathring c=0~.
\eea
Using again that $S$ is a (3,0) and (0,3) form, it is easy to prove that the (4,0)
part of the above equation gives
\bea
dS^{4,0}+ 2 d \mathring c^{4,0}=0~.
\la{u40s}
\eea
Similarly, the (3,1) component gives
\bea
d\mathring c^{3,1}=0~,
\la{u31s}
\eea
 and the (2,2) component implies
\bea
(di_IS)^{2,2}=0~.
\eea
Using that $S=S^{3,0}+ S^{0,3}$, the latter can be expressed as the algebraic condition
\bea
\mathring c^{\bar\g}{}_{\a_1\a_2}S_{\bar\b_1\bar\b_2\bar\g}-\mathring c^\g{}_{\bar\b_1\bar\b_2}S_{\a_1\a_2\g}=0~.
\la{u22s}
\eea
This concludes the analysis of (\ref{eqn2}) for this case.
It is also straightforward to see that (\ref{eqn3}) implies that $dA$ is
a (1,1) form on $M$.

\subsubsection{Geometry}

It is clear that the KY equation (\ref{eqn1}) can be solved for a
family of connections parameterized by a (3,0) and (0,3) form $S$ subject
to the condition that the Nijenhuis tensor of the almost complex structure
is a 3-form (\ref{ac1}). If $S=0$, then there is a unique metric connection $\hat{\mathring \nabla}$ with skew-symmetric
torsion $\mathring c$ such that the almost complex structure is parallel $\hat{\mathring \nabla} I=0$ subject again to the same geometric condition. $\mathring c$
is uniquely determined in terms of the metric and the  Hermitian form $\omega$ \cite{ivanov, uggp}.

The second condition
(\ref{eqn2}) imposes additional restrictions on both $\mathring c$ and $S$.
These are given in (\ref{u40s}), (\ref{u31s}) and (\ref{u22s}). Only (\ref{u40s})
and (\ref{u22s}) restrict $S$.  Both of these have a solution,
\bea
S^{3,0}=-2 \mathring c^{3,0}~.
\la{usols}
\eea
that may not be necessarily unique. It is therefore clear that there are more than one connections with skew-symmetric torsion which solve both
(\ref{eqn1}) and (\ref{eqn2}) for manifolds with an $U(n)$-structure.

\newsection{$SU(n)$ structure}

The fundamental forms are the Hermitian form $\omega$ and the (n,0) form $\epsilon$.
So symmetries are generated\footnote{Symmetries can be generated by either the
 real or imaginary components of $\epsilon$ separately. However, we shall take that both the real and imaginary parts generate symmetries simultaneously.
 If $n>3$, the two different cases give the same conditions.}
 by $\omega$ and  $\epsilon$.
In an adapted basis, we have
 \bea
ds^2=2\delta_{\a\bar\b} e^\a e^{\bar\b}~,~~~\omega=-i \delta_{\a\bar\b} e^\a\wedge e^{\bar\b}~,~~~\epsilon={1\over n!} \epsilon_{\a_1\dots\a_n} e^{\a_1}\wedge\dots \wedge e^{\a_n}~,~~~n\geq3~.
\la{framesun}
\eea
It is apparent that the analysis of the conditions for invariance of the action under the symmetries generated by $\omega$ is  the same as that we have presented for the $U(n)$-structures. So we shall solve
(\ref{eqn1}), (\ref{eqn2}) and (\ref{eqn3}) for the symmetries generated by $\epsilon$.

First let us consider the KY equation (\ref{eqn1}). Up to a complex conjugation, there are four different
arrangements of the indices that this conditions does not vanish identically and so imposes
some restriction on the couplings. Performing the  calculation in the frame adapted
 in (\ref{framesun}), we  find that the (n+1,0) component gives
\bea
\hat\Omega_{\a,\b}{}^\b=0~,
\eea
the $(n,1)$ component gives
\bea
n \hat \Omega_{\bar\b,\a}{}^\a+\hat \Omega_{\a, \bar\b}{}^\a=0~,
\eea
while for two different arrangements of the anti-holomorphic indices the (n-1,2) components give
\bea
&&n\hat\Omega_{\bar\b_1, \bar\a\bar\b_2}+ \hat\Omega_{\bar\b_2,\bar\a\bar\b_1}=0~,
\cr
&&\hat\Omega_{\bar\b_1, \bar\a\bar\b_2}- \hat\Omega_{\bar\b_2,\bar\a\bar\b_1}=0~.
\eea
As we have explained all components of the frame connection $\hat\Omega$
of $\hat\nabla$ that appear in the above expressions belong to $\sigma(\hat\nabla)$.

The above conditions can be solved to express some of the components of the flux
in terms of the geometry and also find the conditions on the geometry implied by (\ref{eqn1}). In particular,
one has
\bea
c_{\b\a}{}^\a=2\Omega_{\b,\a}{}^\a~,~~~c_{\a\b\g}=2\Omega_{[\a,\b\g]}~,
\la{sunc}
\eea
and
\bea
\Omega_{\b,\a}{}^\a=\Omega_{\bar\a, \b}{}^{\bar\a}~,~~~\Omega_{\a,\b\g}=\Omega_{[\a,\b\g]}~.
\la{sungeom}
\eea
In (\ref{sunc}) some of the components of $c$ are expressed in terms of the geometry.
Observe that the (2,1) and (1,2) and traceless component of $c$ is not restricted.

Both  conditions in  (\ref{sungeom}) are restrictions  on the geometry of $M$. The first can be expressed as the
vanishing condition of a linear combination of
the 4th and 5th Gray-Hervella classes \cite{chiossi} of an $SU(n)$
structure while the latter implies that the Nijenhuis tensor of
$M$ is a 3-form. Such conditions have appeared before\footnote{One difference   is
that $M$ is almost complex while in \cite{cardoso} is complex. Observe that
 the class $W_1$  expressed in terms of the Nijenhuis tesnor
  does not vanish.}in the analysis of
geometries with $SU(n)$ structure compatible with a connection with skew-symmetric torsion \cite{cardoso}.
The geometric conditions  (\ref{sungeom})  are significant and imply
that there is a metric connection $\hat{\mathring \nabla}$ with skew-symmetric torsion $\mathring c$
such that
\bea
\hat{\mathring \nabla} \omega=\hat{\mathring \nabla}\epsilon=0~,
\la{sunpar}
\eea
ie the holonomy of $\hat{\mathring \nabla}$ is contained in $SU(n)$. In addition $\mathring c$ is unique and it is determined
 in terms of the metric and the fundamental  form $\omega$ as in the $U(n)$ case, see also \cite{ivanov, uggp}.

To continue observe that the solution of (\ref{eqn1}) for $L=\epsilon$ can be written  as
\bea
c=\mathring c+S~,
\la{csun}
\eea
where now
\bea
S_{\a\b\g}=0~,~~~S_{\b\a}{}^\a=0~,
\la{suns}
\eea
but in general $S^{2,1}\not=0$. $S$ is a 3-form as it is the difference of two
connections with skew-symmetric torsion. It is worth comparing the non-vanishing components
of $S$  with those of the $U(n)$ case in (\ref{uns}).

It remains to solve (\ref{eqn2}). For this, we observe that
\bea
i_\epsilon S=0~,
\eea
as a consequence of (\ref{suns}). Furthermore, we also have as a consequence of
the Bianchi identity that ${\cal L}_\epsilon \mathring c=0$. Thus (\ref{feqn2})
can be rewritten as
\bea
i_\epsilon dS+{n-1\over n+1} i_\epsilon d\mathring c=0~.
\eea
In turn this gives that
\bea
dS^{4,0}+{n-1\over n+1}  d\mathring c^{4,0}&=&0~,
\cr
dS_{\a_1\a_2\b}{}^\b+{n-1\over n+1}  d\mathring c_{\a_1\a_2\b}{}^\b&=&0~.
\la{eqn2sun}
\eea
The first equation can be rewritten as an algebraic equation on $S$ because
$S$ is (2,1) and (1,2) form on $M$. It is not straightforward to solve these
equations for $S$. Nevertheless, they can be easily evaluate them for particular examples.

The third invariance equation (\ref{eqn3}) can be easily solved. It is easy
to see that
\bea
dA_a{}^a=0~.
\eea
The other components are not restricted.

\subsection{Geometry}

\subsubsection{$\epsilon$ symmetries}

The solution to the KY equation for symmetries generated by $\epsilon$ require that $M$ is geometrically restricted by  (\ref{sungeom}). These are  precisely
the conditions for $M$ to admit a metric connection $\hat{\mathring \nabla}$ with
 skew-symmetric torsion $\mathring c$ such that the holonomy of $\hat{\mathring \nabla}$
 is included in $SU(n)$. The torsion $\mathring c$ is determined  in terms
 of the metric and Hermitian form $\omega$. So if the metric and almost
 complex structure are fixed,   $\hat{\mathring \nabla}$
 is unique.  However, the connection $\hat\nabla$ that solves
 the KY equation  is not unique. There is a family of solutions
 parameterized  by the (2,1) and (1,2) and
traceless form $S$ as in (\ref{csun}). This is the full content of the KY
equation.

The second invariance condition (\ref{eqn2}), or equivalently (\ref{feqn2}),
imposes additional conditions. These are given in (\ref{eqn2sun}). If $S=0$,
the resulting  conditions can be viewed as further restrictions   on the geometry as
$\mathring c$ is determined in terms of the metric and Hermitian form. However,
if $S\not=0$, they can be viewed as equations for $S$. It is not apparent that
these determine $S$ uniquely. For example, $S$ is specified up to the exterior
derivative of a co-closed 2-form provided that it is (2,1) and (1,2) form.

\subsubsection{$\omega$ and $\epsilon$ symmetries}

So far we have investigated the conditions for either $\omega$ or $\epsilon$ to generate a symmetry. In the $SU(n)$ case, there is the possibility that both these fundamental forms
generate symmetries. An inspection of the conditions (\ref{uuns}), (\ref{uns}) (\ref{csun}) and (\ref{suns}) reveals that in such case
it is required that
\bea
c=\mathring c~,
\eea
and so $\hat\nabla=\hat{\mathring \nabla}$. As a result, the holonomy of $\hat\nabla$
is contained in $SU(n)$ and both fundamental tensors $\omega$ and $\epsilon$ are parallel.

The second symmetric conditions requires that
\bea
d\mathring c^{4,0}=d\mathring c^{3,1}=0~,
\eea
ie the only non-vanishing component of $d\mathring c$ is the (2,2). This is a condition
on the geometry as $\mathring c$ is expressed in terms of the metric and Hermitian form. It is straightforward to write this condition in terms
of $\omega$ using the expression for $\mathring c$ in \cite{ivanov, uggp}.

\newsection{ $Sp(n)$ and $Sp(n)\cdot Sp(1)$ structures}

\subsection{ $Sp(n)$ structure}

A manifold with a $Sp(n)$ structure admits an almost hyper-complex structure, ie three almost complex structures $(I,J,K)$ satisfying the algebra of imaginary unit quaternions, $I^2=J^2=-1$, $IJ=-JI=K$. The metric $g$ is Hermitian with respect to all three complex structures and so there are associated Hermitian forms $(\omega_I, \omega_J, \omega_K)$.

The three Hermitian forms generate three additional anti-commuting symmetries for the worldline  action (\ref{actn1}). If the complex structures are integrable,  the symmetries satisfy the standard supersymmetry algebra in one dimension. Therefore the action (\ref{actn1})  admits four supersymmetries. If the almost complex structures are not integrable, the additional symmetries are again supersymmetries but now the closure of the  algebra requires the addition  of new generators associated with symmetries generated by the Nijenhuis tensors.

To solve the KY equation (\ref{eqn1}), one can easily adapt the calculation\footnote{We have not given details. The calculation is lengthly but straightforward. The KY equation implies that the  component of the connection $\hat\nabla$  which lies in the complement of the subspace $\mathfrak{sp}(n)$ in the space of 2-forms vanishes. } in \cite{gpp} to reveal that   that all three almost complex structures
$I,J$ and $K$ are parallel with respect to $\hat\nabla$,
\bea
\hat\nabla I=\hat\nabla J=\hat\nabla K=0~.
\eea
Therefore the holonomy of $\hat\nabla$ is in $Sp(n)$ and $M$ is an almost HKT manifold \cite{hkt}.
Fixing the almost hyper-complex structure and the metric, $\hat\nabla$ and so $c$ are unique. The torsion can be expressed in terms of the metric and  Hermitian
forms as in the $U(n)$ case.

The invariance condition (\ref{eqn2}) can be expressed as (\ref{neqn2}) which in turn implies that
\bea
i_I dc=i_J dc=i_K dc=0~,
\eea
ie $dc$ is (2,2) form with respect to all almost complex structures. Similarly,
the third invariance condition (\ref{eqn3}) implies that $dA$ is (1,1) form with respect
to all three almost complex structures and so lies in $\mathfrak{sp}(n)$.

\subsection{$Sp(n)\cdot Sp(1)$ structure}

Manifolds with a $Sp(n)\cdot Sp(1)$ structure admit an almost quaternionic structure which
can be locally represented  by a basis of three almost complex structures $(I,J,K)$
satisfying the algebra of imaginary unit quaternions.
These are compatible with a metric and so there are three associated  Hermitian forms
$(\omega_I, \omega_J, \omega_K)$.  In terms of these, the fundamental form which
generates the symmetry is
\bea
\lambda=\omega_I\wedge \omega_I+\omega_J\wedge \omega_J+\omega_K\wedge \omega_K~.
\eea
Observe that $\lambda$ is a 4-form on $M$ because it is invariant under local $SO(3)$ patching conditions which rotate the
three almost complex structures of the basis.

An investigation reveals that for $n>1$ the KY equation (\ref{eqn1}) implies that $\lambda$ is parallel. The proof is based on a lengthy but straightforward calculation and is similar to that given for the case without torsion in \cite{gpp}.  Therefore, we have that
\bea
\hat\nabla \lambda=0~.
\eea
Thus $M$ is an almost QKT manifold \cite{qkt} and $c$ is uniquely determined in terms of
the fundamental forms and the metric \cite{swann}.

Since the KY equation implies that $\lambda$ is $\hat\nabla$-parallel,
the second invariance equation (\ref{eqn2}) can be reexpressed as (\ref{feqn2})
and so we have that
\bea
i_\lambda dc=0~.
\eea
In addition a direct calculation reveals that the third invariance
condition (\ref{eqn3}) implies that $dA$ lies in $\mathfrak{sp}(n)\oplus
\mathfrak{sp}(1)$.

\newsection{$G_2$ structure}

\subsection{$\varphi$ symmetry}

The fundamental forms are a 3-form $\varphi$ and its dual 4-form $\star\varphi$.
To solve the KY equation (\ref{eqn1}),  we write $\hat\nabla=\pi(\hat\nabla)+\sigma(\hat\nabla)$ following to the decomposition
of 2-forms  $\Lambda^2(\bR^7)=\mathfrak{g}_2+ \Lambda_{\bf 7}$ in $G_2$ representations,
see section 2 where the general procedure is described. We have that $\pi(\hat\nabla)\varphi=0$
and $\sigma(\hat\nabla)$ lies in the 7-dimensional representation $\Lambda_{\bf 7}$. Thus one can set
\bea
\sigma(\hat\nabla)_i{}^j{}_k=L_{im} \varphi^{mj}{}_{k}~,
\eea
for some $L$ tensor.

Suppose now that  $\varphi$ generates a symmetry. The
KY equation (\ref{eqn1}) depends only on the
$\sigma(\hat\nabla)$.   A similar calculation
to that  in \cite{gpp} implies that $L_{ij}=\beta \delta_{ij}$ for some function $\beta$.  Thus we have that$
\hat\nabla=\pi(\hat\nabla)+ \beta \varphi$.

It remains to specify $\pi(\hat\nabla)$. Since  $\sigma(\hat\nabla)=\beta \varphi$
is a 3-form, $\sigma(\hat\nabla)$ is again a metric connection with skew-symmetric torsion which in addition satisfies $\pi(\hat\nabla)\varphi=0$. Such a connection is unique and it exists provided that $G_2$ structure on $M$ is restricted
as
\bea
d\star\varphi= \theta_\varphi \wedge \star\varphi~,
\la{g2geom}
\eea
where $\theta_\varphi$ is the Lee form of $\varphi$.
Following the notation of the previous sections we write $\pi(\hat\nabla)=
\hat{\mathring \nabla}$. The associated 3-form torsion $\mathring c$
is uniquely determined in terms of the metric and $\varphi$, and it is given in \cite{ivanov}, see also \cite{uggp}. Thus there is a family of
solutions
\bea
\hat\nabla=\hat{\mathring \nabla}+ \beta\, \varphi
\eea
to the KY equation (\ref{eqn1})  labeled by $\beta$. 

The second invariance condition (\ref{neqn2}) gives
\bea
-22\,d\beta\wedge \star\varphi+i_\varphi d\mathring c=0~.
\la{gg22eqn2}
\eea
To see this substitute $c=\mathring c+ \beta \varphi$ in (\ref{neqn2}) using ${\cal L}_\varphi \mathring c=0$ and the geometric conditions (\ref{g2geom}) to get
\bea
-22\,d\beta\wedge \star\varphi-6\beta\, \theta_\varphi\wedge \star\varphi+2\beta i_\varphi d\varphi+i_\varphi d\mathring c=0~,
\eea
where $i_\varphi \varphi=-6 \star\varphi$.
Furthermore, observe
that the Kernel of $i_\varphi$ in the space of 4-forms is $\Lambda_{\bf 1}\oplus
\Lambda_{\bf 27}$. Thus $i_\varphi d\varphi$ depends only on  the  $\Lambda_{\bf 7}$
component of $d\varphi$. In particular,
one can show that $i_\varphi d\varphi=3 \theta_\varphi\wedge \star\varphi$, and so
(\ref{eqn2}) gives (\ref{gg22eqn2}).

It remains to solve (\ref{eqn3}). Using the decomposition $\Lambda^2(\bR^7)=\mathfrak{g}_2+ \Lambda_{\bf 7}$, one can easily show that $i_\varphi
dA=0$ implies that $dA$ lies in $\mathfrak{g}_2$.

\subsection{$\star\varphi$ symmetry}

A similar analysis to that presented in the previous section
reveals that the KY equation (\ref{eqn1}) associated with the symmetry generated by  $\star\varphi$
 implies that $\star\varphi$ is $\hat\nabla$-parallel. As a result,
the only solution to the KY equation is
\bea
\hat\nabla=\hat{\mathring\nabla}
\eea
where $\hat{\mathring\nabla}$  is defined as in the previous section. Thus the
solution in this case is unique.

Furthermore, the second invariance condition for $c=\mathring c$ implies that
\bea
i_{\star\varphi} d\mathring c=0~.
\la{g2eqn2}
\eea
Using the decomposition of $\Lambda^4(\bR^7)= \Lambda_{\bf 1}\oplus
\Lambda_{\bf 7}\oplus \Lambda_{\bf 27}$ under $G_2$, one finds that (\ref{g2eqn2}) implies that $d\mathring c$  lies in $\Lambda_{\bf 1}\oplus \Lambda_{\bf 27}$. As $\mathring c$
is determined in terms of the metric and $\varphi$, (\ref{g2eqn2}) becomes a condition on the geometry of $M$. 
Again it is not apparent how to solve it in general. Nevertheless it will be straightforward
to verify it for particular examples. The third invariance condition $i_{\star\varphi} dA=0$ gives that $dA$ lies in $\mathfrak{g}_2$.

\subsection{$\varphi$ and $\star\varphi$ symmetries}

Combining the results of the previous two section, we find that the KY
equation has a unique solution given by the connection with skew-symmetric torsion
and holonomy contained in $G_2$, $\hat\nabla=\hat{\mathring \nabla}$. Furthermore,
the second condition for the invariance of the action (\ref{eqn2}) implies
that $i_\varphi d\mathring c=i_{\star\varphi} d\mathring c=0$. As a consequence,
$d\mathring c$ is restricted to lie in $\Lambda_{\bf 1}\oplus \Lambda_{\bf 27}$.

\newsection{$Spin(7)$ structure}

Every 8-dimensional manifold with a $Spin(7)$ structure admits a unique compatible
connection $\hat{\mathring \nabla}$ with skew-symmetric torsion \cite{ivanov2}. Thus
one can always write $\hat\nabla=\hat{\mathring \nabla}+S$, $\pi(\hat\nabla)=\hat{\mathring \nabla}$ and $\sigma(\hat\nabla)= S$, and investigate
the possibility that there is a $S\not=0$ which solves (\ref{eqn1}). The fundamental
form is a self-dual 4-form $\phi$. Since $\Lambda^2(\bR^8)=\mathfrak{spin}(7)\oplus \Lambda_{\bf 7}$, one takes $S$ to lie in $\bR^8\otimes \Lambda_{\bf 7}$ representation. Moreover since $S$ is the difference between two metric connections with skew-symmetric torsion, $S$ is a 3-form. Using these  a direct substitution in (\ref{eqn1}) and after some calculation\footnote{This calculation is most easily done by writing
the fundamental form $\phi$ in terms of $SU(4)$ fundamental forms.}, one finds that $S=0$.

Since $\hat\nabla=\hat{\mathring \nabla}$, the second invariance condition (\ref{eqn2})
is given in terms of (\ref{neqn2}). So this can be written as
\bea
i_\phi d\mathring c=0~.
\eea
To find the restriction that this equation imposes on $d\mathring c$, one uses
the decomposition $\Lambda^4(\bR^8)=\Lambda_{\bf 1}\oplus \Lambda_{\bf 7}\oplus
\Lambda_{\bf 27}\oplus \Lambda_{\bf 35}$ under $Spin(7)$, where $\Lambda_{\bf 27}$ is the symmetric
traceless   and  $\Lambda_{\bf 35}$ is the 3-form
representation of $SO(7)$, respectively. The above condition implies that the
$\Lambda_{\bf 7}$ component of $d\mathring c$ must vanish.

It remains to investigate the third invariance condition (\ref{eqn3}). It is straightforward to see that $i_\phi dA=0$ implies that $dA$ lies in $\mathfrak
{spin}(7)$.

\vskip 0.5cm
\noindent{\bf Acknowledgements} \vskip 0.1cm
\noindent K Siampos participated in the early stages of this project. GP thanks the Simons Center for hospitality
where parts of this work were done.
GP is partially supported by  the STFC rolling grant ST/G000/395/1.

\end{document}